\documentclass[preprintnumbers,amsmath,amssymb,latexsym,array,enumerate,11pt, letter,superscriptaddress]{revtex4-1}
\usepackage{graphicx}% Include figure files
\usepackage{dcolumn}% Align table columns on decimal point
\usepackage{bm}% bold math
\usepackage{color}
\usepackage{amssymb}
\usepackage{epsfig}
\usepackage{slashed}
\allowdisplaybreaks

\newcommand{\be}{\begin{equation}}
\newcommand{\ee}{\end{equation}}
\newcommand{\beq}{\begin{eqnarray}}
\newcommand{\eeq}{\end{eqnarray}}

\newcommand{\GeV}{{\rm ~GeV }}

\newcommand{\mybar}[1]%
        {\kern  .6pt\overline{\kern - .6pt#1\kern - .6pt}\kern  .6pt}
\begin{document}

\title{An ${\cal O}$(750) GeV Resonance and Inflation}

\author{Yuta Hamada}
\affiliation{ Department of Physics, Kyoto University, Kyoto 606-8502, Japan}
\author{Toshifumi Noumi}
\affiliation{ Jockey Club Institute for Advanced Study, Hong Kong University of Science and Technology, Clear Water Bay, Hong Kong}
\author{Gary Shiu}
\affiliation{ Department of Physics, University of Wisconsin-Madison, Madison, Wisconsin 53706, USA}
\affiliation{ Jockey Club Institute for Advanced Study, Hong Kong University of Science and Technology, Clear
 Water Bay, Hong Kong}
\author{Sichun Sun}
\affiliation{ Jockey Club Institute for Advanced Study, Hong Kong University of Science and Technology, Clear Water Bay, Hong Kong}
\date{\today}

\begin{abstract}
We study the possibility of a heavy scalar or pseudoscalar in TeV-scale beyond the Standard Model scenarios being the inflaton of the early universe in light of the recent  $ \mathcal{O}$(750) GeV diphoton excess at the LHC. We consider a scenario in which the new scalar or pseudoscalar couples to the Standard Model gauge bosons at the loop level through new massive Standard Model charged vectorlike fermions with or without dark fermions. We calculate the renormalization group running of both the Standard Model and the new scalar couplings, and present two different models that are perturbative and with a stabilized vacuum up to near the Planck scale. Thus, the Standard Model Higgs and this possible new resonance may still preserve the minimalist features of Higgs inflation.
\end{abstract}

\maketitle
\section{Introduction}

One of the recurrent lessons in physics is that simplicity beats complexity.
While theoretical reasonings (such as the hierarchy problem and naturalness) and observations (e.g., dark matter)
 have motivated many beyond the Standard Model (BSM) scenarios, the search for new physics  has so far been elusive. One may wonder if we have seen the beginning of a desert in that the new physics that completes the Standard Model (SM) only appears at higher energies currently not within reach.

Very recently the CMS and ATLAS Collaboration have reported a
roughly $3\sigma$  enhancement in the $\gamma\gamma$ spectrum at $ \mathcal{O}$(750) GeV,
based on
the first $\sim$ 3/fb of data from run 2 of the LHC at $\sqrt{s}=13$ TeV  \cite{CMS:2015dxe,atlas2015}.  
 This potential new resonance can be interpreted as either a spin-zero or a spin-two particle, as the Laudau-Yang theorem states that an on-shell spin-one particle cannot decay to diphoton. A spin-two particle would yield an interesting $\gamma\gamma$ distribution that is peaked in the beam directions. The possibility that the candidate resonance is a KK graviton is not excluded but is quite constrained by the run 1 data. For concreteness, we focus on the interpretation of this candidate resonance as a spin-zero particle, which can be a scalar or a pseudoscalar.
  The 8 TeV run of the LHC did not report a signal, which hints at  the possibility that the production rate might have a steeper energy dependence. All these hints, taken together, suggest that this new resonance couples to the SM gauge bosons including photons and gluons only through higher dimensional operators. The resonance can be produced by $gg \rightarrow \Phi \rightarrow \gamma\gamma$ though $q \overline{q}$ annihilation can also contribute. Moreover, the ATLAS Collaboration has enough events to show a larger width of the resonance, which suggests the presence of some other decay channels besides the diphoton. Since there are no excesses shown in the other channels, one tempting explanation is that this new resonance decays into the dark sector. 
  
This possible excess has already sparked numerous theoretical explorations \cite{Ellis:2015oso,diphoton,Dhuria:2015ufo,Zhang:2015uuo,Salvio:2015jgu}.
To explain this potential new resonance, two prevailing ideas are as follows:
(1) to stay within weakly coupled theories and introduce  vectorlike SM charged fermions, or (2) to invoke strong dynamics as in composite Higgs models. While we maintain our healthy skepticism towards the reality of this signal, we think it worthwhile to explore the cosmological implications of this potential new resonance while awaiting further data from the LHC to either confirm or falsify its existence.  Though previous works \cite{Ellis:2015oso,diphoton,Dhuria:2015ufo,Zhang:2015uuo,Salvio:2015jgu} provide various explanations for the possible excess, the $X(750)$ resonance and the additional new particles are introduced without necessarily having a purpose. Here, we would like to see,  if we take Occam's razor seriously, 
 what correlated statements in cosmology can we make.

Inflation is the leading paradigm of early universe cosmology. 
An inflationary universe can be realized with a scalar or pseudoscalar (known as the inflaton) with a sufficiently flat potential. Moreover, this inflaton should couple to the SM fields somehow in order for reheating to occur. It is in this minimalist spirit that Higgs inflation \cite{Bezrukov:2007ep}, in which the SM Higgs field was identified with the inflation (with non-minimal coupling to gravity), was proposed. Here, we examine how the new resonance, if confirmed, may alter this minimalist scenario.
We focus on the weak coupling explanation of the resonance, as it is challenging (if at all possible)
to accommodate inflation concretely with 
strongly coupled theories (see e.g., Refs.\cite{Anguelova:2010qh,Channuie:2011rq,Jiang:2015qor}). Moreover, a strongly coupled sector typically comes with a plethora of particles, far from the minimalist approach we opt to take.
In the weakly coupled scenarios, the new resonance can be a candidate for the inflaton, while the additional vectorlike SM charged fermions are necessarily there for the inflaton to reheat the SM sector. In a sense, our scenario is a minimal model to accommodate  the success of both the SM and cosmic inflation, if the $X(750)$ resonance is confirmed. 
 We studied the perturbativity and the stability of the SM vacuum in light of the new particles and the associated new couplings.

The paper is organized as follows. We discuss some general aspects of this new resonance in Sec. II. We present some models with or without dark fermions and solve the corresponding renormalization group (RG) equations in Sec. III. The parameter spaces of the initial conditions that preserve perturbativity and vacuum stability of the models are presented. We conclude in Sec. IV. For completeness, 
we also include in the appendix the complete RG equations for the models explaining the $X(750)$ resonance considered in \cite{Ellis:2015oso}.

\section{General aspects of the new resonance}

The CMS and ATLAS excess suggests a cross section:
\beq \label{Eq:sigma_gammagamma}
\sigma (pp \rightarrow \Phi) \times BR( \Phi \rightarrow \gamma\gamma) \gtrsim 2 fb
\eeq
with the mass $M_\Phi \approx 750$ GeV of a new spin-zero particle $\Phi$. As discussed in the Introduction, we consider the five-dimensional operators with a SM singlet scalar or pseudoscalar that couples to the SM gauge bosons. For the scalar case, $\Phi=S$, the effective couplings are given by
 \beq
 L_\text{effective}\supset \frac{S}{4} (-g_{s\gamma } F_{\mu\nu}F^{\mu\nu} -g_{s\gamma } G_{\mu\nu}G^{\mu\nu})\,,
 \eeq
whereas for the pseudoscalar case, $\Phi=P$, we have
 \beq
 L_\text{effective}\supset \frac{P}{4} (-g_{p\gamma } F_{\mu\nu}\tilde{F}^{\mu\nu} -g_{p\gamma } G_{\mu\nu}\tilde{G}^{\mu\nu})\,.
 \eeq
 Here $G_{\mu\nu}$ and $F_{\mu\nu}$ are the Standard Model color $SU(3)_C$ and $U(1)_{em}$ field strength. The gluon couplings account for the gluon-gluon fusion production of this new resonance, and the photon couplings give rise to the dominant decay channels to diphoton. 
Notice here there are also likely decay modes of S/P into dibosons ($\gamma Z$, ZZ, WW) because the dimensional-5 operators with these gauge fields are also allowed. Currently an excess does not show up in the other channels; therefore they are constrained by the run 1 and 2 data \footnote{The diboson resonance ($WW,ZZ,WZ$) around 2 TeV reported earlier this year \cite{Aad:2015owa} by ATLAS might have some correlation to the current excess in the diphoton channel. See, e.g., Refs.\cite{Allanach:2015hba,Allanach:2015gkd} for the physics argument for the earlier diboson excess.}.

 It is well known that the new vectorlike fermion at  the TeV scale can give rise to these effective operators. In fact, such fermions also appear in BSM extensions that address other pressing questions like flavor physics or dark matter issues  \cite{Sun:2013cza, Alves:2015mua}.
 Let us  introduce here $N_\psi$ vectorlike fermions $\psi=(\psi_L, \psi_R)^T$ in the $3$- representation of the color $SU(3)_C$ and with an electromagnetic charge $Q$.
Their Yukawa couplings to $S$ and the mass term are
 \beq
 L \supset -\lambda_{S\psi\psi} S \bar{\psi}\psi  - M_\psi \bar{\psi}\psi \,.
 \eeq
Here and in what follows we focus on the scalar case, $\Phi=S$, for concreteness, but the results for the pseudoscalar case should be qualitatively the same. We also assume suppressed Yukawa couplings between vectorlike quarks and the SM fermions.
By integrating out the heavy fermion we arrive at the effective couplings:
 \begin{align}
& g_{sg}= \frac{N_\psi \lambda_{S\psi\psi} \alpha_s}{3\pi \sqrt{2}M_\psi}\,,\\
& g_{s\gamma}= \frac{\sqrt{2}N_\psi \lambda_{S\psi\psi} Q^2 \alpha}{\pi M_\psi}\,.
\end{align}
The LHC measurement production times  branching ratio is $\sigma (pp \rightarrow \Phi) \times BR( \Phi \rightarrow \gamma\gamma)  \propto g_{sg}^2 g_{s\gamma}^2/ (8g_{sg}^2+ g_{s\gamma}^2)$, with $g_{sg}/g_{s\gamma} \gg 1$. Taking $\alpha_s(\text{TeV}) \sim 0.09$ from PDG \cite{Agashe:2014kda}, we have
\beq \label{Eq:fitting_condition}
 \left(\frac{2 N_\psi \lambda_{S\psi\psi} Q^2 \alpha}{\pi}{1\text{TeV}\over M_\psi}\right)^2 \gtrsim \frac{2}{13000}\,.
 \eeq

  \section{Specific models of the new resonance for perturbativity and vacuum stabilization}
  \label{Sec:Models}
  
   In this section we explore models which are perturbative and stable up to near the Planck scale, $10^{17}$GeV, while at the same time explaining the diphoton excess. As we have discussed,
   the diphoton excess motivates
    the existence of a new spin-zero particle and new vectorlike massive fermions with  SM charges. The couplings of this sector at the electroweak scale is constrained by Eq.~\eqref{Eq:fitting_condition}.  In addition, we here would like to include dark fermions which have no SM charges, suggested by a possibility that the dark matter sector couples to the new spin $0$ particle. Our matter content then includes the SM particles, the new scalar $S$, the new SM charged vector-like fermions, and the dark fermions.
    
Let us summarize our notation for the couplings. First, the scalar potential is given by~\footnote{This simple scalar extension of the SM has also appeared in the study of current and future collider probes to new physics and dark matter portals, see e.g. \cite{Craig:2014lda}.}
  \beq
  V=-\mu^2|H|^2+\lambda(|H|^2)^2+m^2 S^2+{\lambda_S\over 4!}S^4+{\kappa\over2}|H|^2S^2\,.
  \eeq 
We introduce $N_\psi$ vectorlike fermions $\psi$ and $N_D$ dark fermions $D$~\footnote{We would like to thank A.Cohen for pointing out that an $S|H|^2$ term can also be present, unless one considers a $Z_2$ symmetry acting both on $S$ and fermions. However, $S|H|^2$ is a dimension-3 soft operator which does not contribute much to the RG running in the UV limit.}. We assume that the vectorlike fermions are in the $3$-representation of $SU(3)_C$, and have an $SU(2)_L$ isospin $S_\psi$ and a hypercharge $Y$. On the other hand, the dark fermions have no SM charges. Their Yukawa couplings to $S$ are denoted by $\lambda_{S\psi\psi}$ and $\lambda_{SDD}$, respectively. More explicitly, $L \supset -\lambda_{S\psi\psi} S \bar{\psi}\psi - \lambda_{SDD} S \bar{D}D $. Their RG equations are then given by
\small{     \begin{align}
  16\pi^2{d\lambda_S\over dt}&=
  					     	3\lambda_S^2+12\kappa^2+24N_\psi(2S_\psi+1)\lambda_{S\psi\psi}^2\lambda_\psi-144(2S_\psi+1)N_\psi\lambda_{S\psi\psi}^4+8N_D\lambda_{SDD}^2\lambda_S-48N_D\lambda_{SDD}^4\,,
						\nonumber \\
  16\pi^2{d\lambda_{S\psi\psi}\over dt}&=
  					     	(3+6N_\psi(2S_\psi+1))\lambda_{S\psi\psi}^3-8g_3^2\lambda_{S\psi\psi}-6Y^2g_Y^2\lambda_{S\psi\psi}-6S_\psi(S_\psi+1)g_2^2\lambda_{S\psi\psi}+2N_D\lambda_{SDD}^2\lambda_{S\psi\psi}\,,
						\nonumber \\	
  16\pi^2{d\lambda_{SDD}\over dt}&=
  					     	(3+2N_D)\lambda_{S\psi\psi}^3+6N_T(2S_\psi+1)\lambda_{SDD}\lambda_{S\psi\psi}^2\,,
						\nonumber \\		
  16\pi^2{dg_3\over dt}&=
  					  \text{(SM part)}+{2\over3}N_\psi(2S_\psi+1)g_3^3\,,
						\nonumber \\
  16\pi^2{dg_2\over dt}&=
  					  \text{(SM part)}+4N_\psi {S_\psi(S_\psi+1)(2S_\psi+1)\over3}g_2^3 \,,
						\nonumber \\
  16\pi^2{dg_Y\over dt}&=
  					  \text{(SM part)}+4Y^2N_\psi(2S_\psi+1)g_Y^3\,,
						\nonumber \\	
  16\pi^2{dy_t\over dt}&=
  					  \text{(SM part)}\,,
						\nonumber \\	
  16\pi^2{d\lambda\over dt}&=
  					  \text{(SM part)}
					  +{1\over2}\kappa^2\,,
					  \nonumber \\
					  \label{RGeqs}
  16\pi^2{d\kappa\over dt}&=
  					\kappa\left(4\kappa+12\lambda+\lambda_S+6y_t^2+12N_\psi(2S_\psi+1)\lambda_{S\psi\psi}^2+4N_D\lambda_{SDD}^2-{3\over2}g_Y^2-{9\over2}g_2^2\right)\,,	  
\end{align}}
where $g_3$, $g_2$, and $g_Y$ are the gauge couplings for $SU(3)_C$, $SU(2)_L$, and $U(1)_Y$, respectively. $t=\ln \mu$ is the renormalization scale, $y_t$ is the top Yukawa coupling and $\lambda$ is the Higgs self-coupling. The calculation here is based on \cite{Cheng:1973nv}. The SM part and initial conditions can be found in, e.g. Ref. \cite{Buttazzo:2013uya}. In the rest of this section, based on the above RG equations, we investigate two types of models with (a) toplike vectorlike fermions and dark fermions (b) vector-like fermions with exotic SM charges.
 
\medskip
\noindent  {\bf (a) Toplike vectorlike fermions and dark fermions}:

Let us start with a model with $N_T$ vectorlike fermions $T$ with (3,1,2/3) under $SU(3)_C\times SU(2)_L\times U(1)_Y$ and $N_D$ dark Dirac fermions $D$. This class of models is given by $\psi=T$, $S_T=0$, and $Y=2/3$ in the previous RG equations. Note that these dark fermions can be the dark matter of the universe. They can annihilate mainly through $DD \rightarrow S/P \rightarrow gg$ in the early universe. Because in our case the dark fermion masses are not constrained much from perturbativity and vacuum stabilization (as long as they are not far from the ${\cal O}$(TeV) range), we can easily achieve the correct relic abundance. 

%%%%%%%%%%%%%%%%%%%%%
\begin{figure}
\hfill
\includegraphics[width=4cm]{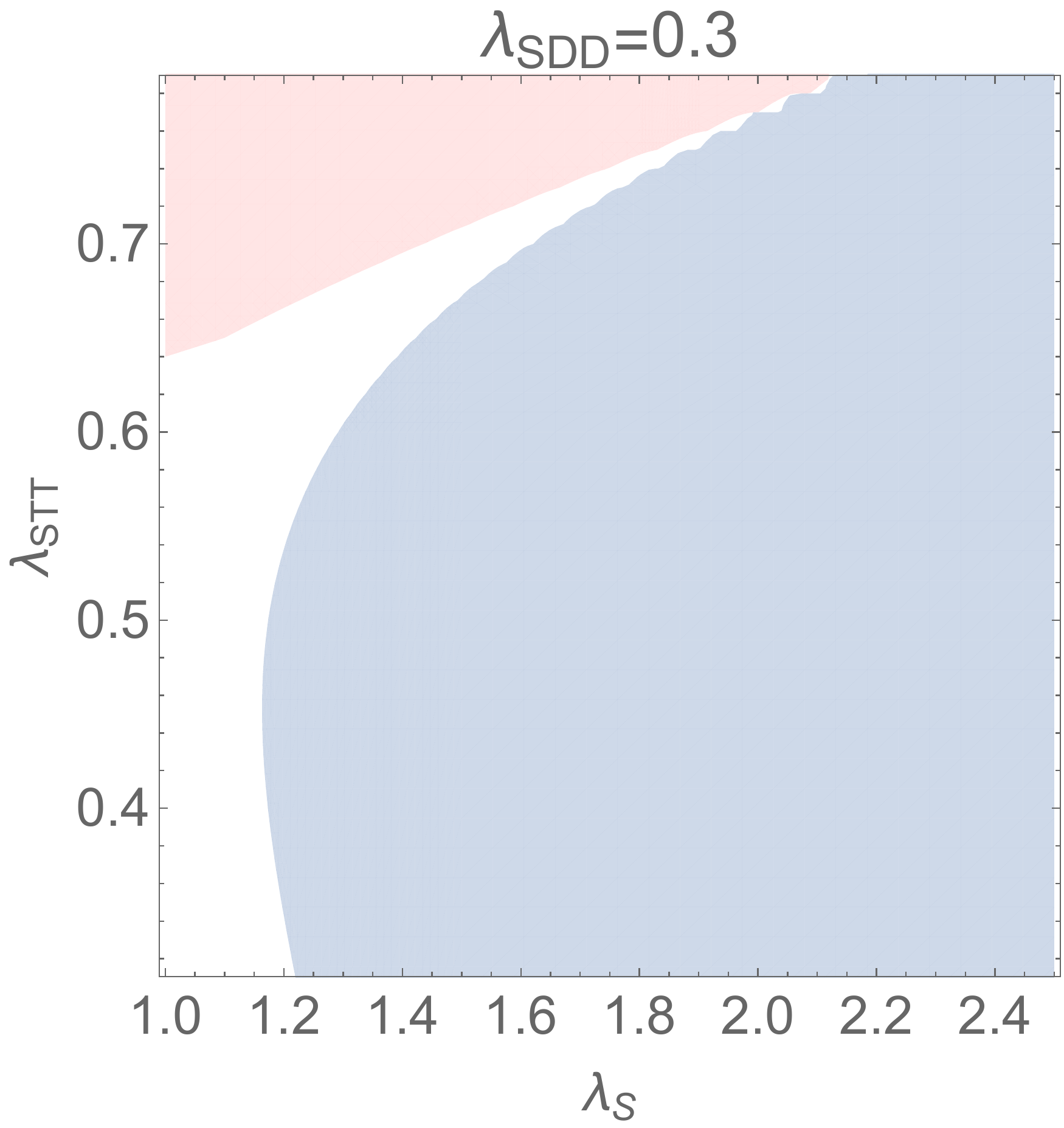}
\hfill
\includegraphics[width=4cm]{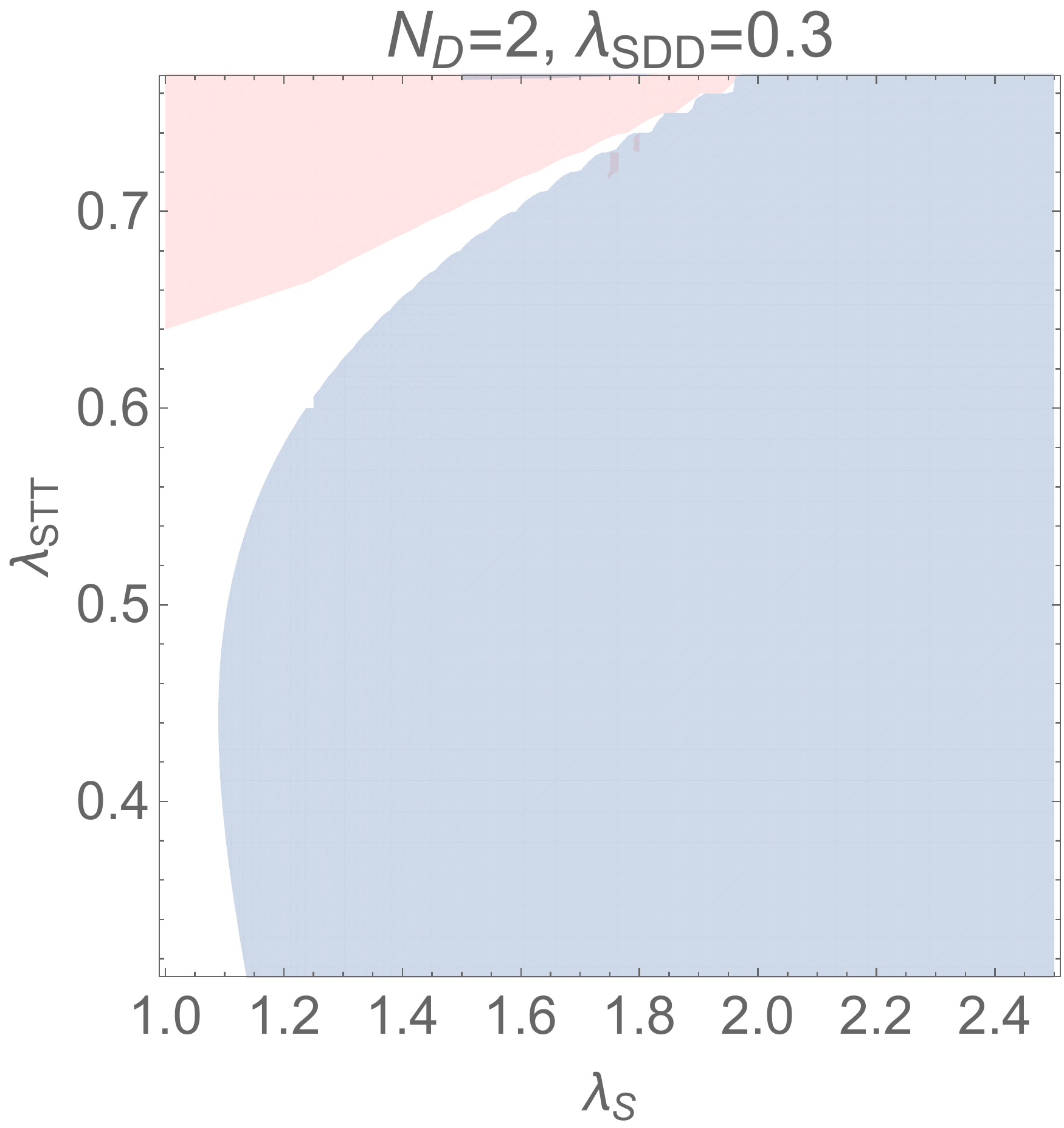}
\hfill
\mbox{}
\caption{ {The blue and red regions are not perturbative and stable up to $10^{17}$GeV, respectively. In the left(right) panel, we take $M_T=1000$GeV, $\lambda_{SDD}=0.3, N_D=1(2)$ and $N_T=1$.}}
\label{DF}
\end{figure}
%%%%%%%%%%%%%%%%%%%%%

%%%%%%%%%%%%%%%%%%%%%
\begin{figure}
\hfill
\includegraphics[width=4cm]{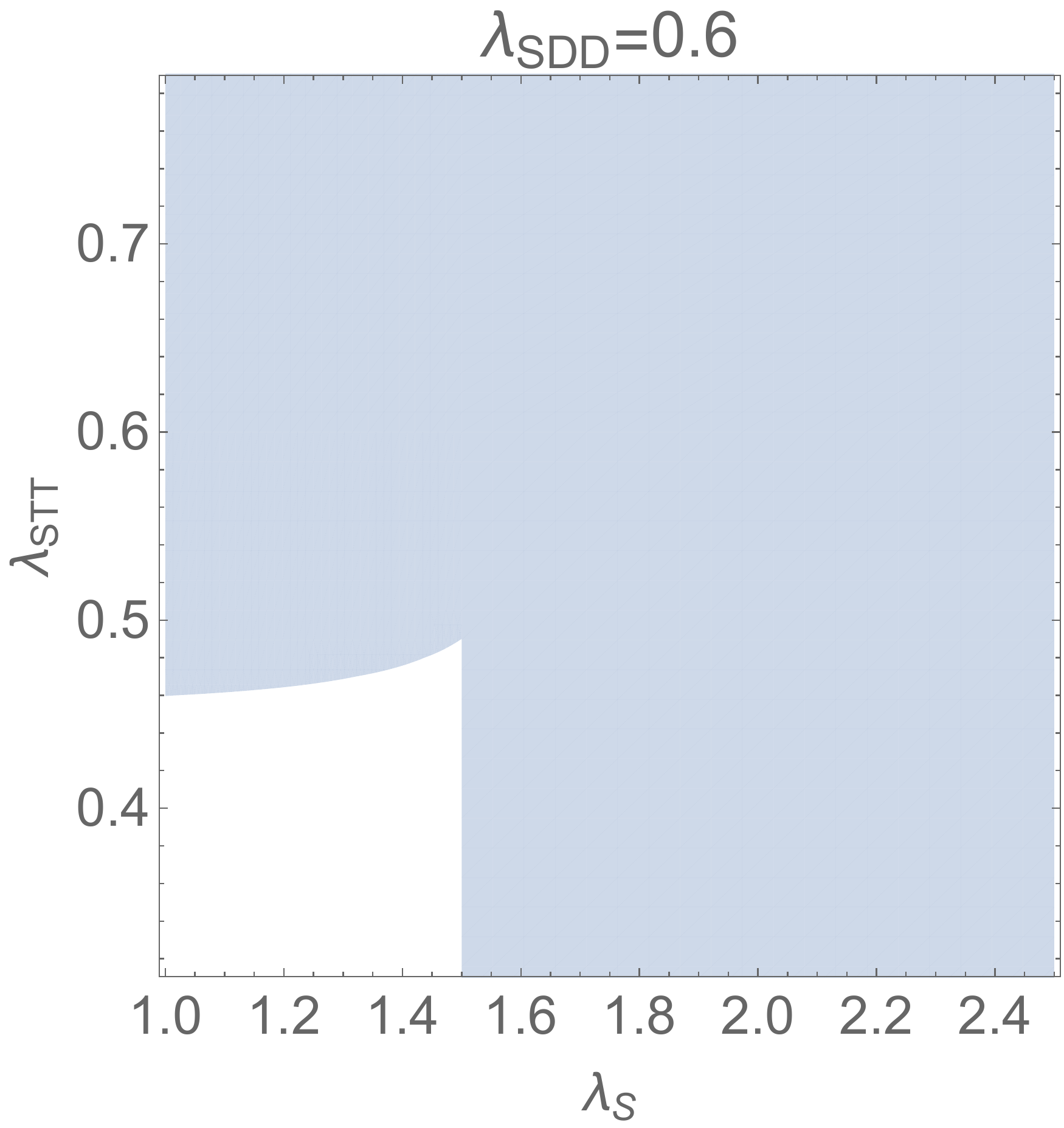}
\hfill
\includegraphics[width=4cm]{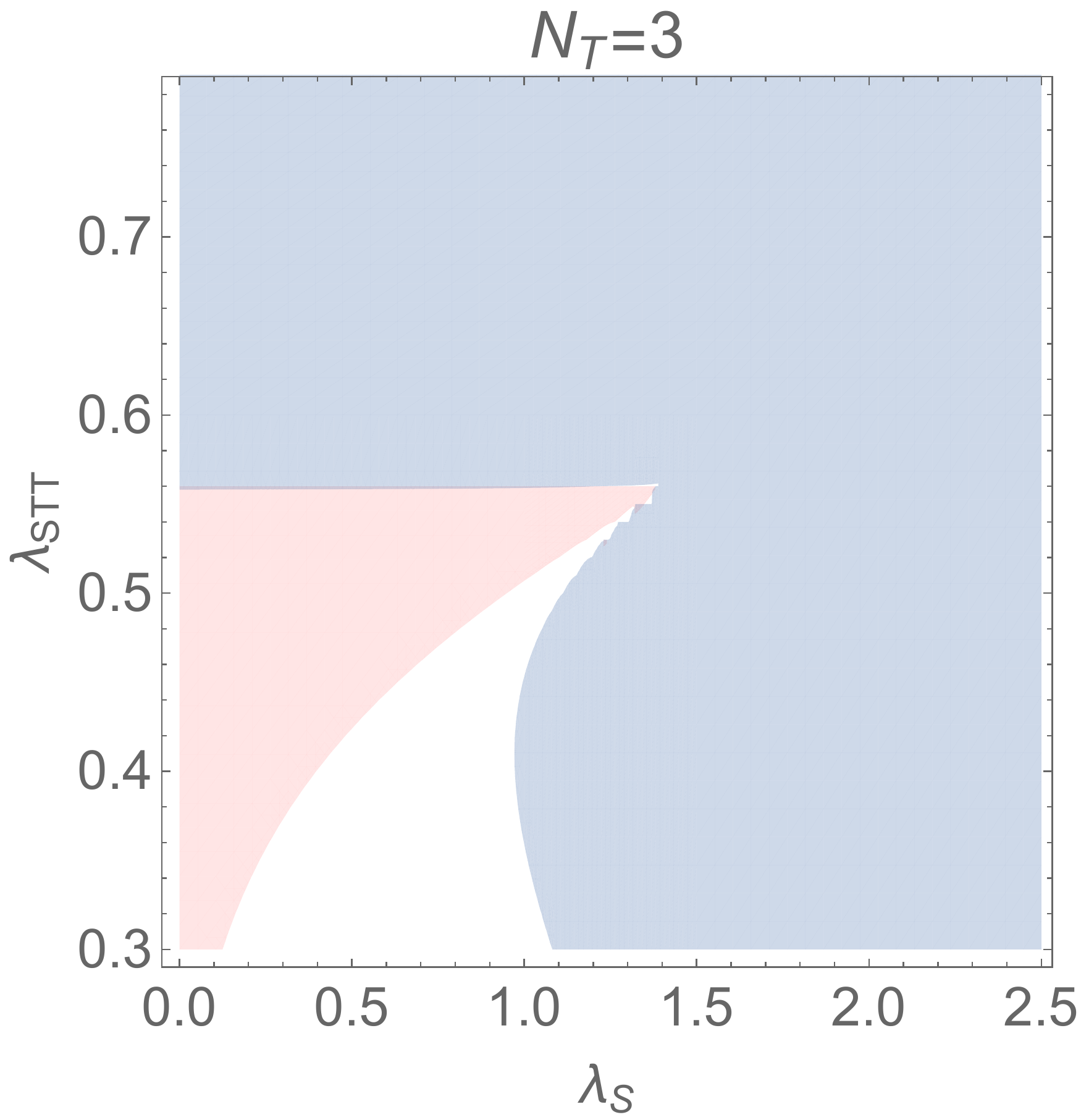}
\hfill
\mbox{}
\caption{ { 
Left panel: Same as Fig.~\ref{DF}, but taking $\lambda_{SDD}=0.6$.
Right panel: we take $N_T=3$ and $\lambda_{SDD}=0$.
}}
\label{DF2}
\end{figure}
%%%%%%%%%%%%%%%%%%%%%
In Figs.~\ref{DF} and ~\ref{DF2}, we show the region where this model becomes perturbative and stable up to $10^{17}$GeV for $N_T=1$.
The blue and red regions correspond to nonperturbative and unstable regions, respectively.
Unfortunately, we cannot find a region realizing Eq.~\eqref{Eq:fitting_condition}, which requires $\lambda_{S\psi\psi}\gtrsim\mathcal{O}(1)$.
Increasing $\lambda_{SDD}$ and $N_D$ makes the problem worse as in the right panel of Fig.~\ref{DF} and the left panel of Fig.~\ref{DF2}. The dark sector can therefore be an obstruction to realizing a ``desert" up to high scale generically.
On the other hand, if we increase the number of $N_T$, the situation becomes better, because the constraint~\eqref{Eq:fitting_condition} on the Yukawa coupling $\lambda_{STT}$ at the electroweak scale is relaxed.
However, it is not sufficient to explain the diphoton signal, see the right panel of Fig.~\ref{DF2}.

\medskip
\noindent {\bf (b) Exotic charge fermion}:\\ 
  The difficulty of the previous model comes from the necessity of large coupling for one toplike vectorlike fermion.
  From Eq.~\eqref{Eq:fitting_condition}, we see that we can take smaller $\lambda_S$ by considering vectorlike fermion with a larger charge.
  However, if the representation is too large, the gauge coupling hits the Landau pole far below the Planck scale.
   By solving the one-loop renormalization group equation of the gauge couplings, the position of the pole is       
  %----------------------------------------
\begin{align}
\Lambda_{g_i}
&=M_\psi \exp 
\left( {8\pi^2\over b_i}{1\over g_i^2(M_\psi)} \right) \nonumber \\
&=
M_\psi \left( {M_t\over M_\psi} \right)^{b_{i,\text{SM}}/b_i}
\exp \left( {8\pi^2\over b_i}{1\over g_i^2(M_t)} \right),
\end{align}
%----------------------------------------
where $i=Y, 2$, and $b_{i,\text{SM}}$ and $b_i$ are the coefficients of the beta function of $g_i$ in the SM and SM with an exotic quark, respectively.      
We summarize the position of the pole in Table~\ref{Table:pole}, from which we can see that only $Q\leq 5/3$ is consistent with perturbativity up to $10^{17}$GeV. The position of the Landau pole of $\lambda_{S\psi\psi}$, $\Lambda_{\lambda_{S\psi\psi}}$, is also shown. We note that $\Lambda_{\lambda_{S\psi\psi}}$ is independent of $\kappa$ and $\lambda_{S}$.
        %++++++++++++++++++++++Table:pole++++++++++++++++++++++++++++
\begin{table}[t]
\begin{center}
\begin{tabular}{|c|c|c|c|c|c|}
\hline
$(SU(3)_C,SU(2)_L,U(1)_Y)$ & $Q_\text{max} $ & $\Lambda_{g_Y}\, [\GeV]$&$\Lambda_{g_2}$\, [\GeV]& $\sigma\times BR>2\text{fb}$&$\Lambda_{\lambda_{S\psi\psi}}<10^{17}\text{GeV}$\\ \hline
$(3,5,0)$ & $2$ & $8.9\times 10^{40}$ & $1.3\times 10^{5}$&$\lambda_{S\psi\psi}>0.18\left({M_\psi\over700\text{GeV}}\right)$&--\\ \hline
$(3,4,1/2)$ & $2$ & $7.6\times 10^{26}$ & $6.1\times 10^{7}$&$\lambda_{S\psi\psi}>0.32\left({M_\psi\over700\text{GeV}}\right)$&--\\ \hline
$(3,3,1)$ & $2$ & $4.7\times 10^{16}$ & $1.1\times 10^{20}$&$\lambda_{S\psi\psi}>0.35\left({M_\psi\over700\text{GeV}}\right)$&--\\ \hline
$(3,2,3/2)$ & $2$ & $2.1\times 10^{13}$ & --&$\lambda_{S\psi\psi}>0.35\left({M_\psi\over700\text{GeV}}\right)$&--\\ \hline
$(3,1,2)$ & $2$ & $1.8\times 10^{14}$ & --&$\lambda_{S\psi\psi}>0.44\left({M_\psi\over700\text{GeV}}\right)$&--\\ \hline
$(3,3,2/3)$ & $5/3$ & $1.8\times 10^{24}$ & $1.1\times 10^{20}$&$\lambda_{S\psi\psi}>0.53\left({M_\psi\over700\text{GeV}}\right)$&$\lambda_{S\psi\psi}<0.75$\\ \hline
$(3,2,7/6)$ & $5/3$ & $3.4\times 10^{17}$ & --&$\lambda_{S\psi\psi}>0.54\left({M_\psi\over700\text{GeV}}\right)$&$\lambda_{S\psi\psi}<0.81$\\ \hline
$(3,1,5/3)$ & $5/3$ & $2.3\times 10^{17}$ & --&$\lambda_{S\psi\psi}>0.62\left({M_\psi\over700\text{GeV}}\right)$&$\lambda_{S\psi\psi}<1.02$\\ \hline
\end{tabular}
\caption{
The possible $SU(2)_L\times U(1)_Y$ charge.
$Q_\text{max}$ represents the maximum electromagnetic charge among the multiplets.
Only the particle with $Q_\text{max}\leq 5/3$ can be perturbative up to $10^{17}$GeV. %
Here we take $M_\psi=700$GeV.
}
\label{Table:pole}
\end{center}
\end{table}
%++++++++++++++++++++++Table:pole++++++++++++++++++++++++++++
Therefore, as simple successful examples, we consider vectorlike fermions whose charges are (3,2,7/6) and (3,1,5/3) under $SU(3)_C\times SU(2)_L\times U(1)_Y$, and show that these models indeed explain the diphoton excess and can be valid up to near the Planck scale.
The mass bound on a charge $5/3$ particle is about $900\GeV$~\cite{CMS:2015alb}, if $\psi$ mainly decays into top, while the bound is $690$GeV if the main decay mode is a light quark~\cite{Aad:2015tba}.
%%%%%%%%%%%%%%%%%%%%%
\begin{figure}
\hfill
\includegraphics[width=4cm]{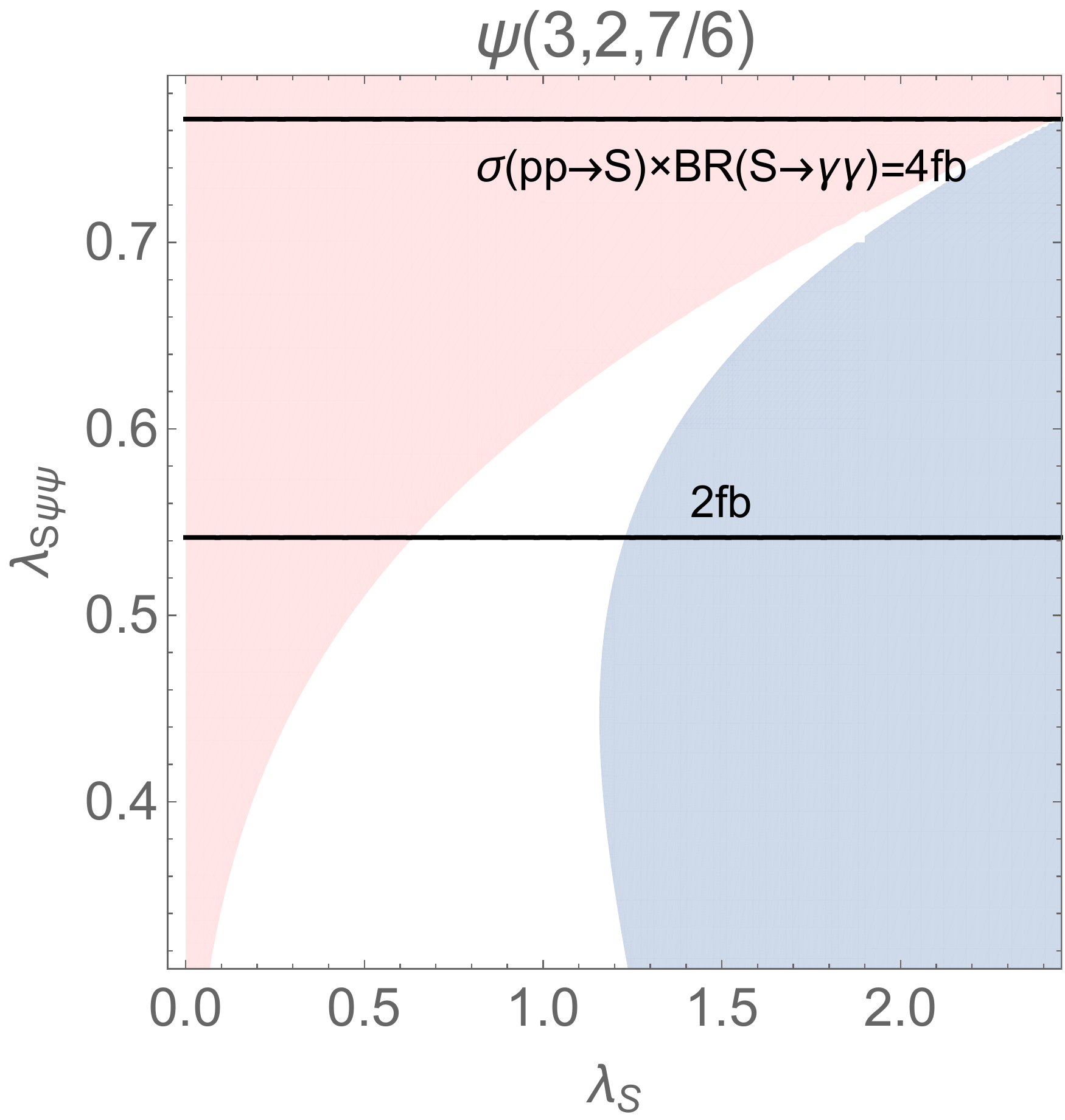}
\hfill
\includegraphics[width=5cm]{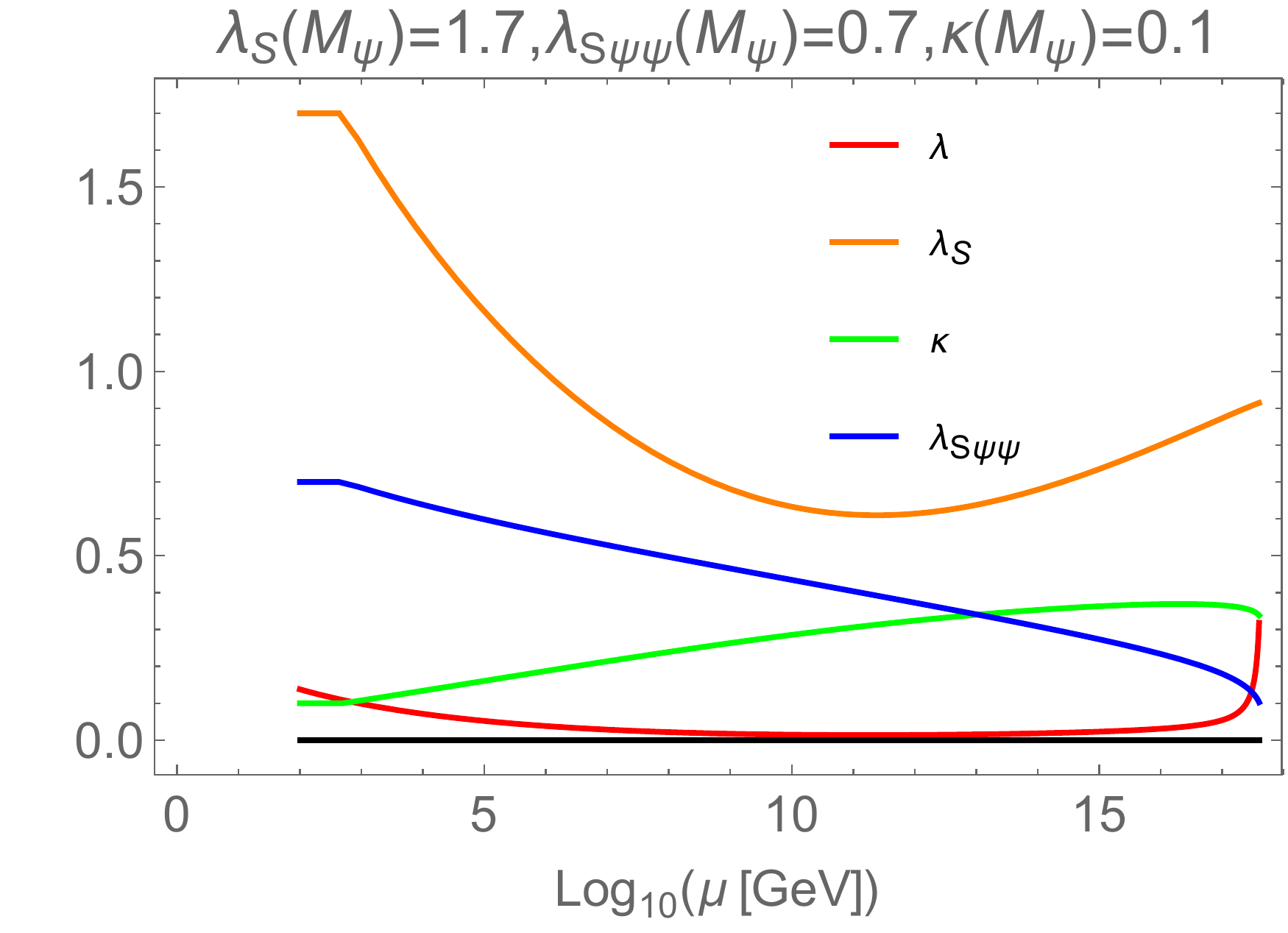}
\hfill
\mbox{}
\caption{ { 
Left panel: The blue and red regions are not perturbative and stable up to $10^{17}$GeV, respectively.
The white region is allowed. We take $\kappa=0$ and $M_\psi=700$GeV.
Right panel: The running of scalar and Yukawa couplings. All couplings are perturbative and stable up to $10^{17}$GeV.
}}
\label{exotic}
\end{figure}
%%%%%%%%%%%%%%%%%%%%%

%%%%%%%%%%%%%%%%%%%%%
\begin{figure}
\includegraphics[width=5cm]{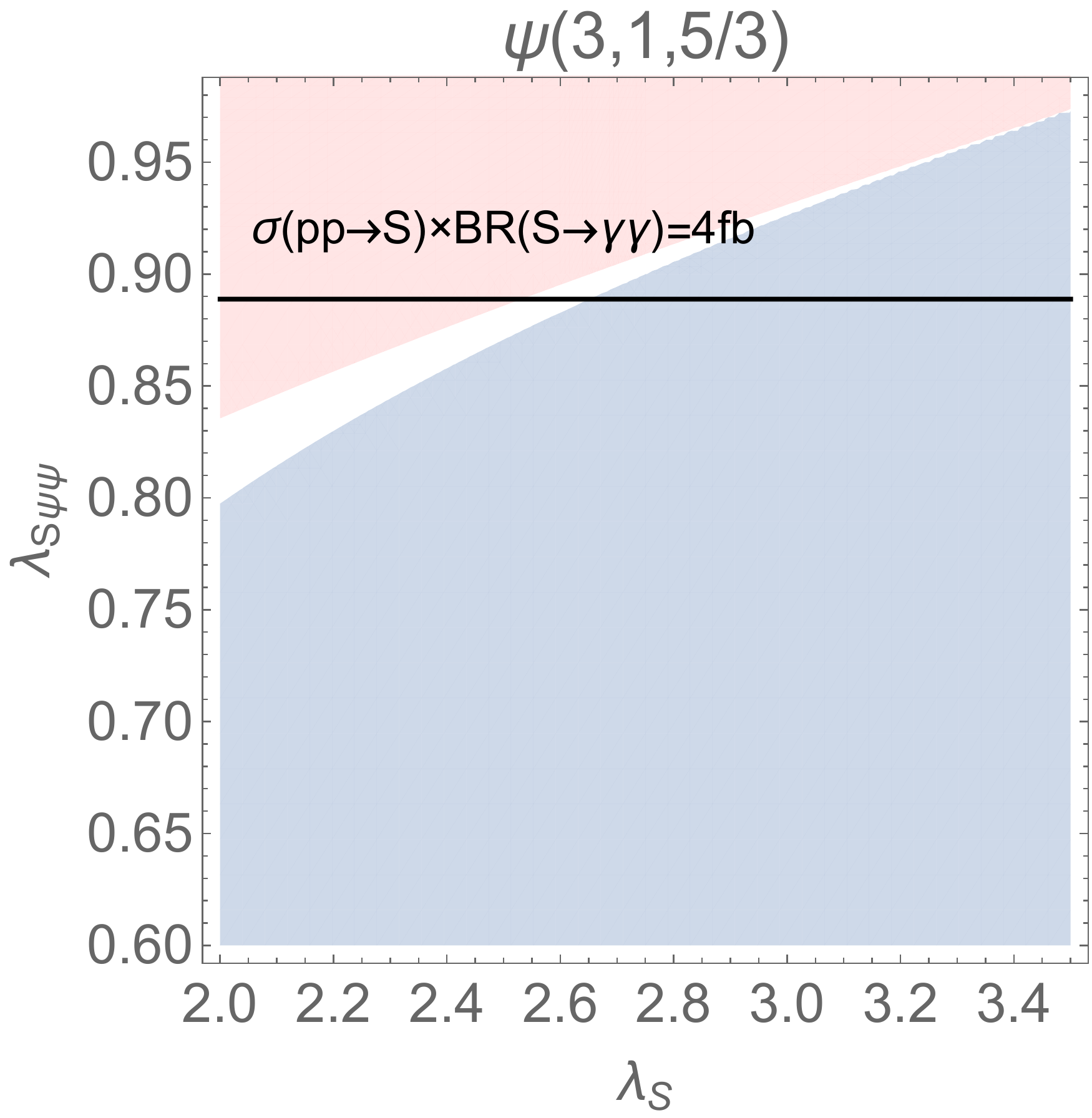}
\caption{ { 
Same as Fig.~\ref{exotic}, but $S_\psi=0$, $Y=5/3$.
}}
\label{exotic2}
\end{figure}
%%%%%%%%%%%%%%%%%%%%%

%
We plot the region excluded by perturbativity and stability up to $10^{17}$GeV, in Fig.~\ref{exotic} with $M_\psi=700$GeV, $S_\psi=2$ and $Y=7/6$.
The inclusion of $\kappa$ does not change the allowed region very much.
It is found that only a very small part in the white region has $\sigma(pp\to S)\times BR(S\to\gamma\gamma)\sim2\text{fb}$.
In Fig.~\ref{exotic2}, we also show the viable region with $S_\psi=0, Y=5/3$.
Although the region is small, the consistent region survives.
\medskip

  \section{Conclusion}
  
Motivated by the recent $\mathcal{O}(750)$ GeV diphoton excess at the LHC, we have studied the RG running in the scenario where a new spin $0$ particle couples to the SM gauge bosons at the loop level through new massive SM charged vectorlike fermions. For both models with and without dark fermions, we have explicitly shown that there exists a certain parameter range which accommodates the stability and perturbativity up to $10^{17}$ GeV. We illustrated two types of workable models with vectorlike fermions with exotic SM charges which may explain the current diphoton excess at the same time. Our result thus suggests a possibility that, other than the X(750) resonance and the associated vectorlike fermion in the ${\cal O}$(TeV) range, there is a ``desert" up to near the Planck scale.

In the same spirit of Higgs inflation, it is natural to look for the inflaton among the two (pseudo)scalars, i.e., the SM Higgs and the new spin-zero particle, of the minimal setups we proposed. In the standard Higgs inflation, the action for the inflaton $\varphi$ in the Jordan frame is given by
\begin{align}
S_J\simeq\int d^4x \left[-\frac{M_{\rm Pl}^2}{2}R-\xi \varphi^2 R+\frac{1}{2}(\partial \varphi)^2-\frac{\lambda_{\rm inf}}{4}\varphi^4\right]\,,
\end{align}
which is characterized by the inflaton quartic coupling $\lambda_{\rm inf}$ and the nonminimal coupling $\xi$ to the scalar curvature. In the region $|\xi|\gg1$, however, the action in the Einstein frame is reduced to~\footnote{
As pointed out in~\cite{Burgess:2009ea,Bezrukov:2010jz}, the cutoff scale derived from the tree-level unitarity is generically around the background inflaton value during inflation. Although it might suggest that nonrenormalizable operators could spoil the flatness of the inflationary potential (a common feature for large field inflation models in general), we assume, as in the conventional Higgs inflation scenarios, that the inflaton potential is protected by a scale symmetry in the Jordan frame which is required by the unknown UV complete theory. See also a recent paper~\cite{McDonald:2016cdh} for a possibility that the existence of a new scalar may help push up the cutoff scale and resolve this issue.}
\begin{align}
S_E\simeq\int d^4x \left[-\frac{M_{\rm Pl}^2}{2}R+\frac{1}{2}(\partial \chi)^2-\frac{M_{\rm Pl}^2}{4}\frac{\lambda_{\rm inf}}{\xi^2}\left(1+\exp\left(-\frac{2\chi}{\sqrt6 M_{\rm Pl}}\right)\right)\right]\,,
\end{align}
which depends only on the ratio $\lambda_{\rm inf}/\xi^2$. It can directly be related to the length of the inflationary period, so that we have, e.g., $\xi=4.5\times10^{4}\sqrt{\lambda_{\rm inf}}$ for 60 e-folds.
Since both the SM Higgs and the new spin-zero particle may have the quartic coupling of the order $\lesssim\mathcal{O}(1)$, either can be used as the inflaton $\varphi$ by introducing an appropriate size of the nonminimal coupling $\xi$. The primordial tilt $n_s$ and the tensor-to-scalar ratio $r$ are the same as the standard Higgs inflation, $n_s\simeq 0.97$ and $r\simeq0.0033$ for the e-folding number $N=60$, so they agree well with the current constraints by Planck~\cite{Ade:2015xua} and BICEP2/Keck Array~\cite{Array:2015xqh}. This is the conclusion of this paper.

One important question to be investigated next is whether the inflaton is the SM Higgs or the new spin-zero particle. The main differences between those particles are their couplings to the SM particles and the dark fermions (if they exist). Such a difference may leave some interesting imprints, e.g., on the reheating process and the dark matter abundance.  Moreover, both scalar fields may play the role of the inflaton and thus our setups naturally suggest a realization of multifield inflation. It would be interesting to explore this possibility. The isocurvature remnants would be a key feature of this class of models and would be useful to constrain details of the model such as the reheating process and the initial condition of the inflationary trajectory~\cite{progress}\footnote{Notice that if only one of the two scalars  plays the role of the inflaton, the inflationary dynamics is reduced to the single field one and no significant isocurvature remnants are generated during inflation.}.
Another interesting direction would be to embed inflation into other scenarios explaining the current diphoton excess, such as models in which the new spin-zero particle is identified with a dilaton or an axion. To explain the diphoton excess, the decay constant of the spin-zero particle is of the order of the electroweak scale.
At least naively, the dilaton would go into the strong coupling regime at the inflation scale. An electroweak scale decay constant 
is also not suited for the simplest form of axion inflation.
Therefore, it is not easy to realize inflationary models such as dilaton inflation or  axion inflation that explains the diphoton excess simultaneously. However, it would be interesting to explore this direction further.
We hope to address these issues elsewhere.

\bigskip
\noindent \begin{acknowledgments}
Y.H. thanks Hikaru Kawai and Koji Tsumura for useful discussions.
Y.H. is supported in part by the Grant-in-Aid for JSPS Fellows No.25$\cdot$1107.
T.N, G.S. and S.S. are supported in part by the CRF Grants of the Government of the Hong Kong SAR under HUKST4/CRF/13G
and the HKGRC Grants No. 604213 and No. 16304414.
G.S. is supported in part by the DOE grant DE-FG-02-95ER40896. G.S. thanks the University of Chinese Academy of Sciences for hospitality when this work is completed. 
 \end{acknowledgments}

\bigskip
\noindent\textbf{Note added:} While this work was being completed, other papers \cite{Dhuria:2015ufo,Zhang:2015uuo,Salvio:2015jgu} on electroweak vacuum stabilization appeared on the arXiv. Our analysis concerns more general setups and the implications to inflation.  We also like to also point out that the Refs. \cite{Dhuria:2015ufo,Zhang:2015uuo} did not take into account the Landau pole of gauge couplings $g_Y$ and $g_2$, which is appropriately taken into account in our analysis. See Table I. The model in \cite{Salvio:2015jgu} contains vectorlike fermions with the electromagnetic charge $3/2$, which may form rather exotic mesons with a fractional electromagnetic charge, whereas the electromagnetic charges of mesons in our models are always integers.

\appendix

  \section{renormalization group equations for various models in \cite{Ellis:2015oso}}

For completeness, we here provide the RG equations for various models introduced in~\cite{Ellis:2015oso}. The models I and II there may be embedded into our model in Sec.~\ref{Sec:Models}, whose RG equations are given by Eq.~\eqref{RGeqs}. The RG equations for the models III and IV are given below.
  
{\small  
\medskip

 \noindent    {\bf MODEL III~\cite{Ellis:2015oso}}:
       \begin{align}
 16\pi^2{d\lambda_\Phi\over dt}&=
  					     	3\lambda_{\Phi}^2+12\kappa^2-144 \lambda_{SBB}^4+24 \lambda_{SBB}^2 \lambda_{\Phi} 
						 -288 \lambda_{SQQ}^4+48 \lambda_{SQQ}^2 \lambda_{\Phi} -144\lambda_{STT}^4+24 \lambda_{STT}^2 \lambda_{\Phi},					
						\nonumber \\
  16\pi^2{d\lambda_{STT}\over dt}&=
  					     	6 \lambda_{SBB}^2 \lambda_{STT}
				+12 \lambda_{SQQ}^2 \lambda_{STT}+9 \lambda_{STT}^3-8g_3^2\lambda_{STT}-{8\over3}g_Y^2\lambda_{STT},
						\nonumber \\							
  16\pi^2{d\lambda_{SQQ}\over dt}&=
  					     	6 \lambda_{SBB}^2 \lambda_{SQQ}
   						+15 \lambda_{SQQ}^3+6 \lambda_{SQQ} \lambda_{STT}^2-8g_3^2\lambda_{SQQ}-{9\over2}g_2^2\lambda_{SQQ}-{1\over6}g_Y^2\lambda_{SQQ},
						\nonumber \\							
  16\pi^2{d\lambda_{SBB}\over dt}&=
  					     	9 \lambda_{SBB}^3+12\lambda_{SBB} \lambda_{SQQ}^2
						+6 \lambda_{SBB} \lambda_{STT}^2-8g_3^2\lambda_{SBB}-{2\over3}g_Y^2\lambda_{SBB},							
						\nonumber \\	
  16\pi^2{dg_3\over dt}&=
  					  \text{(SM part)}+{2\over3}g_3^3+{4\over3}g_3^3+{2\over3}g_3^3,
						\nonumber \\
  16\pi^2{dg_2\over dt}&=
  					  \text{(SM part)}+2g_2^3,
						\nonumber \\
  16\pi^2{dg_Y\over dt}&=
  					  \text{(SM part)}+{16\over9}g_Y^3+{1\over9}g_Y^3+{2\over9}g_Y^3,
						\nonumber \\	
  16\pi^2{dy_t\over dt}&=
  					  \text{(SM part)}
						\nonumber \\							
  16\pi^2{d\lambda\over dt}&=
  					  \text{(SM part)}
					  +{1\over2}\kappa^2,	
					  \nonumber \\
  16\pi^2{d\kappa\over dt}&=
  					\kappa\left(4\kappa+12\lambda+\lambda_\Phi+6y_t^2+12\lambda_{STT}^2+24\lambda_{SQQ}^2+12\lambda_{SBB}^2-{3\over2}g_Y^2-{9\over2}g_2^2\right).	  								
  \end{align}

\medskip
\noindent
     {\bf MODEL IV~\cite{Ellis:2015oso}}:
       \begin{align}
  16\pi^2{d\lambda_\Phi\over dt}&=
  					     	3\lambda_{\Phi}^2+12\kappa^2-144 \lambda_{SBB}^4+24 \lambda_{SBB}^2 \lambda_{\Phi} -48 \lambda_{SEE}^4+8 \lambda_{SEE}^2 \lambda_{\Phi} 
						\nonumber\\
						&-96 \lambda_{SLL}^4+16\lambda_{SLL}^2 \lambda_{\Phi} -48 \lambda_{SNN}^4+8 \lambda_{SNN}^2\lambda_{\Phi} -288 \lambda_{SQQ}^4+48 \lambda_{SQQ}^2 \lambda_{\Phi} -144\lambda_{STT}^4+24 \lambda_{STT}^2 \lambda_{\Phi},					
						\nonumber \\
  16\pi^2{d\lambda_{STT}\over dt}&=
  					     	6 \lambda_{SBB}^2 \lambda_{STT}+2 \lambda_{SEE}^2 \lambda_{STT}+4 \lambda_{SLL}^2 \lambda_{STT}+2 \lambda_{SNN}^2\lambda_{STT}
						\nonumber \\
   						&+12 \lambda_{SQQ}^2 \lambda_{STT}+9 \lambda_{STT}^3-8g_3^2\lambda_{STT}-{8\over3}g_Y^2\lambda_{STT},
						\nonumber \\							
  16\pi^2{d\lambda_{SQQ}\over dt}&=
  					     	6 \lambda_{SBB}^2 \lambda_{SQQ}+2 \lambda_{SEE}^2 \lambda_{SQQ}+4 \lambda_{SLL}^2 \lambda_{SQQ}+2 \lambda_{SNN}^2\lambda_{SQQ}
						\nonumber \\
   						&+15 \lambda_{SQQ}^3+6 \lambda_{SQQ} \lambda_{STT}^2-8g_3^2\lambda_{SQQ}-{9\over2}g_2^2\lambda_{SQQ}-{1\over6}g_Y^2\lambda_{SQQ},
						\nonumber \\							
  16\pi^2{d\lambda_{SBB}\over dt}&=
  					     	9 \lambda_{SBB}^3+2 \lambda_{SBB} \lambda_{SEE}^2+4 \lambda_{SBB} \lambda_{SLL}^2+2 \lambda_{SBB} \lambda_{SNN}^2+12\lambda_{SBB} \lambda_{SQQ}^2
						\nonumber\\
						&+6 \lambda_{SBB} \lambda_{STT}^2-8g_3^2\lambda_{SBB}-{2\over3}g_Y^2\lambda_{SBB},							
						\nonumber \\	
  16\pi^2{d\lambda_{SLL}\over dt}&=
  					     	6 \lambda_{SBB}^2 \lambda_{SLL}+2 \lambda_{SEE}^2 \lambda_{SLL}+7 \lambda_{SLL}^3+2 \lambda_{SLL} \lambda_{SNN}^2+12\lambda_{SLL} \lambda_{SQQ}^2
   						\nonumber \\
   						&+6 \lambda_{SLL} \lambda_{STT}^2-{9\over2}g_2^2\lambda_{SLL}-{3\over2}g_Y^2\lambda_{SLL},
						\nonumber \\							
  16\pi^2{d\lambda_{SEE}\over dt}&=
  					     	6 \lambda_{SBB}^2 \lambda_{SEE}+5 \lambda_{SEE}^3+4 \lambda_{SEE} \lambda_{SLL}^2+2 \lambda_{SEE} \lambda_{SNN}^2+12\lambda_{SEE} \lambda_{SQQ}^2
						+6 \lambda_{SEE} \lambda_{STT}^2-6g_Y^2\lambda_{SEE},										
						\nonumber \\	
  16\pi^2{d\lambda_{SNN}\over dt}&=
  					     6 \lambda_{SBB}^2 \lambda_{SNN}+2 \lambda_{SEE}^2 \lambda_{SNN}+4 \lambda_{SLL}^2 \lambda_{SNN}+5 \lambda_{SNN}^3+12
   \lambda_{SNN} \lambda_{SQQ}^2+6 \lambda_{SNN} \lambda_{STT}^2,							
						\nonumber \\													
  16\pi^2{dg_3\over dt}&=
  					  \text{(SM part)}+{2\over3}g_3^3+{4\over3}g_3^3+{2\over3}g_3^3,
						\nonumber \\
  16\pi^2{dg_2\over dt}&=
  					  \text{(SM part)}+2g_2^3+{2\over3}g_2^3,
						\nonumber \\
  16\pi^2{dg_Y\over dt}&=
  					  \text{(SM part)}+{16\over9}g_Y^3+{1\over9}g_Y^3+{2\over9}g_Y^3+{1\over3}g_Y^3+{2\over3}g_Y^3,
						\nonumber \\	
  16\pi^2{dy_t\over dt}&=
  					  \text{(SM part)}
						\nonumber \\							
  16\pi^2{d\lambda\over dt}&=
  					  \text{(SM part)}
					  +{1\over2}\kappa^2,	
					  \nonumber \\
  16\pi^2{d\kappa\over dt}&=
  					\kappa\left(4\kappa+12\lambda+\lambda_\Phi+6y_t^2+12\lambda_{STT}^2+24\lambda_{SQQ}^2+12\lambda_{SBB}^2+8\lambda_{SLL}^2+4\lambda_{SEE}^2+4\lambda_{SNN}^2-{3\over2}g_Y^2-{9\over2}g_2^2\right).	  			
  \end{align}

}

 \end{document}